\documentclass[twocolumn,secnumarabic,amssymb, amsmath, nofootinbib,tightenlines,
nobibnotes, aps, prl,epsfig]{revtex4}
\usepackage{graphicx}
\usepackage{dcolumn}
\usepackage{bm}
\begin{document}
\preprint{APS/123-QED}
\title{Nuclear longitudinal structure function in eA processes at the LHeC}

\author{G.R.Boroun}%
 \email{grboroun@gmail.com; boroun@razi.ac.ir }
\author{B.Rezaei }
\altaffiliation{brezaei@razi.ac.ir}
\author{S.Heidari }
\affiliation{ Physics Department, Razi University, Kermanshah
67149, Iran}
\date{\today}
\begin{abstract}
The nucleon and nuclear longitudinal structure functions are
determined by the Kharzeev-Levin-Nardin (KLN) model of the low $x$
gluon distribution. The behavior of the gluon distribution ratio
$R_{g}={G^{A}}/{AG^{p}}$ and the ratio
 $R_{L}^{total}={F_{L}^{A-total}}/{AF_{L}^{p-total}}$ in this processes are found.
 The heavy longitudinal structure function ratios  in eA processes at the LHeC region are discussed.
Heavy contributions to the ratio of the  total longitudinal
structure function are considerable and should not be neglected
especially at smaller $x$ of the LHeC project.
 In the KLN model
  the new
 geometrical scaling for transition from the linear to nonlinear regions
 in accordance with  the LHeC processes is used, whose results intensively depended on  the heavy quarks mass effect.\\

\end{abstract}
 \pacs{***}
\maketitle
\subsection{I. Introduction}

 Our knowledge of the gluon distribution function of free nucleons
 comes from the deep inelastic scattering (DIS ) measurements in
 lepton-nucleon collisions, as at low-$x$ the gluon distributions
 are predominant at all values of $Q^{2}$. There is a transition from the linear to nonlinear
 regions as it can be tamed by screening effects. These
 nonlinear terms reduce the growth of the gluon distribution at
 low $x$ values. Therefore DIS processes in LHeC provide very
 important tools for probing the gluon distribution in
 the nucleons and  nuclei. The nuclear gluon distribution
 $xg^{A}(x,Q^{2})$ can be determined from the gluon distribution of
 nucleons which are bounded in a nucleus. Also, the nuclear distribution
 functions can be extracted from the measurements of deep
 inelastic lepton-nucleus scattering (eA processes). In the
 electron-proton/ion collider LHeC, we intended to demonstrate how
 the low $x$ data are possible for nuclear targets and could
 constrain the nuclear gluon distribution function.\\
 The LHeC shows an increase
in the kinematic range of the deep inelastic scattering (DIS)
because the DIS kinematics are $2 < Q^{2}< 100,000~GeV^{2}$ and
$0.000002 < x < 0.8$ with a center-of-mass energy of about
$\sqrt{s_{ep}}>1~TeV$. Clearly this
 increase in the precision of parton
distribution functions (PDF$^{,}$s) at low-$x$ kinematic region is
expected to  cause by the non-linear effects in the so-called saturation region [1-4].\\
The nuclear parton distribution functions (nPDF$^{,}$s) can be
determined based on the DGLAP [5-6] evolution, analogous to the
parton distributions of the free proton. At low $x$, the data show
a reduction of the nuclear distribution functions with respect to
the free distribution functions. This phenomenon is caused by the
nuclear shadowing effects as $xg^{A}(x,Q^{2})<Axg^{N}(x,Q^{2})$.
These shadowing corrections give rise to nonlinear terms in the
evolution equation of the gluon distribution function. Indeed,
these behaviors are tamed by the saturation effects. In the gluon
saturation approach, an important point is the $x$-dependence
saturation scale $Q_{s}^{2}(x)$ where it is the critical line
between the linear and nonlinear regions. It is expected that the
nonlinear effects are small in $Q^{2}>Q^{2}_{s}(x)$ and
 it should be strong in $Q^{2}<Q^{2}_{s}(x)$ which generate
geometrical scaling in this region. Therefore the nuclear reduced
cross section is dependent upon the single variable
$\tau_{0}=\frac{Q^{2}}{Q_{s}^{2}(x)}$, as $\sigma_{
\gamma^{*}A}(x,Q^{2})=\sigma_{ \gamma^{*}A}(\tau_{0})$ and the
saturation scale is given by
$Q_{s}^{2}(x)=Q_{0}^{2}(\frac{x}{x_{0}})^{-\lambda}$ [7-9]. Here
$Q_{0}^{2}=0.34~GeV^{2}$, $x_{0}=3.0{\times}10^{-3}$ and exponent
$\lambda$ is a dynamical
quantity at the order of $\lambda{\simeq}0.25$.\\
The paper is organized as follows: Section II deals with KLN model
in transition from traditional geometrical scaling to new
geometrical scaling with respect to heavy quarks mass. Section III
introduces the nuclear longitudinal structure function by
considering the heavy quarks portion. In section IV we finalized
the total longitudinal structure function calculation for light
and
heavy nuclei. Finally, in Sec.V we summarized the results.\\

\subsection{II. KLN model and saturation scale}

We focused on the nuclear longitudinal structure function based on
the geometrical scaling at low $x$. The main purpose of this study
was to analyze possible compatibility of the new geometrical
scaling with the KLN model for transition from the linear to
nonlinear nuclear behavior. In heavy production, the geometrical
scaling is expected to be violated by heavy quarks mass, since the
traditional geometrical scaling ($\tau_{0}$) can be modified to
take into account heavy quarks mass [9]:
\begin{eqnarray}
\tau_{H}=(1+\frac{4m_{H}^{2}}{Q^{2}})^{1+\lambda}\frac{Q^{2}}{Q^{2}_{0}}(\frac{x}{x_{0}})^{\lambda}.
\end{eqnarray}
In the KLN model [10], a simple relation for the unintegrated
gluon distribution was observed as it is related to the gluon
distribution by the following form
\begin{eqnarray}
G(x,Q^{2})=\int^{Q^{2}}dk_{t}^{2}\varphi(x,k_{t}^{2}),
\end{eqnarray}
where $\varphi$ is the unintergrated gluon distribution of a
nucleon or nucleus. Authors in Ref.[10] used a simplified
assumption about the form of $G(x,Q^{2})$ by two regions of
integration over $Q^{2}$ defined in accordance with the critical
line $Q_{s}^{2}$. For a nucleon we  use the KLN Ansatz as
\begin{eqnarray}
G(x,Q^{2})=\bigg{\{}^{\frac{K_{0}S}{\alpha_{s}(Q_{s}^{2})}Q^{2}(1-x)^{4}~~,~~~Q^{2}<Q_{s}^{2}(\tau_{0}<1)}
_{\frac{K_{0}S}{\alpha_{s}(Q_{s}^{2})}Q_{s}^{2}(1-x)^{4}~~,~~~Q^{2}>Q_{s}^{2}(\tau_{0}>1),}
\end{eqnarray}
where the numerical coefficient $K_{0}$ can be determined from the
gluon density, which is usually taken from the parameterization
groups and $S$ is the area corresponds to the target.
Here the factor $(1-x)^{4}$ is to describe the fact that the gluon density is small at high $x$ values.\\
 The gluon distribution function, for a nucleus with the mass number $A$, can be exploited to $G^{A}(x,Q^{2})$ with
replacements $S{\rightarrow}S^{A}=A^{\frac{2}{3}}S$ and
$Q_{s}^{2}{\rightarrow}Q_{s}^{2A}=A^{\frac{1}{3}}Q_{s}^{2}$ [11].
The gluon distribution for a nucleus with respect to  Eq.3 can be
written as
\begin{eqnarray}
G^{A}(x,Q^{2})=\bigg{\{}^{\frac{K_{0}S^{A}}{\alpha_{s}(Q_{s}^{2A})}Q^{2}(1-x)^{4}~~,~~~Q^{2}<Q_{s}^{2A}(\tau^{A}_{0}<1)}
_{\frac{K_{0}S^{A}}{\alpha_{s}(Q_{s}^{2A})}Q_{s}^{2A}(1-x)^{4}~~,~~~Q^{2}>Q_{s}^{2A}(\tau^{A}_{0}>1).}
\end{eqnarray}
The transition point between the linear and nonlinear regions in
accordance with the critical line for the proton ($A=1$) and a
nucleus ($A$) is
\begin{eqnarray}
x_{c}=x_{0}(\frac{Q^{2}_{0}A^{1/3}}{Q^{2}})^{1/\lambda}.
\end{eqnarray}
Table.1 shows that the critical point is dependent on the mass
number $A$ and $Q^{2}$ values. At low $x$ values, the transition
point between the linear and nonlinear behavior is observable for
light nuclei at low $Q^{2}$ and for heavy nuclei at low and
moderate $Q^{2}$ values in $eA$ processes. These critical points
refer to zero quark mass so that the geometrical scaling defined
by $\tau^{A}_{0}$($=\frac{Q^{2}}{Q_{s}^{2A}(x)}$). Since masse of
heavy quarks in the LHeC region is  not negligible, therefore the
geometrical scaling is sensitive to the mass of the heavy quarks.
This is consistent with a new geometrical scaling $\tau^{A}_{H}$
into the one as follows
$\tau^{A}_{H}=(1+\frac{4m_{H}^{2}}{Q^{2}})^{1+\lambda}\frac{Q^{2}}{Q_{s}^{2A}}$.
We expect that the transition points shift to lower $x$ values as
transition between the linear and nonlinear behaviors tamed at low
$x$ values. As to the mass correction, the critical point is given
by the following form
\begin{eqnarray}
x_{c}=x_{0}(\frac{Q^{2}_{0}A^{1/3}}{Q^{2}(1+\frac{4m_{H}^{2}}{Q^{2}})})^{1/\lambda}.
\end{eqnarray}
In Tables 2-4, we observe the heavy quark mass effects on the
critical points. Indeed, the difference between the (traditional)
geometrical scaling variable ($\tau_0$) and the "new" geometrical
scaling variable ($\tau_H$), particularly for $eA$ ($\tau^{A}_{0}$
and $\tau^{A}_{H}$) is strong dependence on the heavy quark mass
as $\frac{\Delta \tau}{\tau_{0}}$ or $\frac{\Delta
\tau^{A}}{\tau_{0}^{A}}=(1+\frac{4m_{H}^{2}}{Q^{2}})^{1+\lambda}-1$.\\
 Fig.1
shows the behavior of the ratio
$R_{g}=\frac{G^{A}(x,Q^{2})}{AG^{p}(x,Q^{2})}$ for light and heavy
nuclei $A=12$ and $A=208$ respectively at $Q^{2}$ values of
$2,~10$ and $100~GeV^{2}$ in order to determine the gluon
densities in nuclei. The magnitude of shadowing effects are
considered by the geometrical scaling behavior at low $x$ values.
We observe the saturation effects for the ratio $R_{g}$ at
$x<10^{-2}$ and for small values of $Q^{2}$ at light and heavy
nuclei by using the traditional scaling. In Figs.2-3, we observe
that the critical points related to the new geometrical scaling
are noticeable  in a wide region of $x$ ($x<10^{-5}$ upto
$x<10^{-15}$). These observations are essential in determining of
these ratios when we take into
account the heavy quarks mass effects.\\
\subsection{III. Nuclear  longitudinal structure function }

Now, we consider the nuclear longitudinal structure function at eA
processes with respect to the nuclear gluon density behavior. The
nuclear longitudinal structure function is interested because it
is directly sensitive to the nuclear gluon density through the
transition $g^{A}{\rightarrow}q^{A}\overline{q}^{A}$ in eA-DIS.
 Indeed a measurement of $F_{L}^{A}(x,Q^{2})$ can be used to
extract the nuclear gluon structure function. Therefore the
measurement of $F^{A}_{L}$ provides a sensitive test of
perturbative QCD (pQCD). Since the longitudinal structure function
$F_{L}$ contains rather large heavy flavor contributions at
small-x region, so the measurements of these observables in the eA
processes have told us about the ratio of the heavy quarks
contribution to the nuclear longitudinal structure function and
also the dependence of nuclear parton distribution functions
(nPDFs) on heavy quarks mass. In perturbative QCD, the nuclear
longitudinal structure function can be written as
\begin{eqnarray}
x^{-1}F_{L}^{A}&=&C_{L,ns}{\otimes}q_{ns}^{A}+<e^{2}>(C_{L,q}{\otimes}q_{s}^{A}+C_{L,g}{\otimes}g^{A})\nonumber\\
&&+x^{-1}F^{Heavy-A}_{L},
\end{eqnarray}
where the symbol ${\otimes}$ indicates convolution over the
variable $x$ as:
$A(x){\otimes}B(x)=\int_{x}^{1}\frac{dy}{y}A(y)B(\frac{x}{y})$.
Here $q_{ns}^{A}$, $q_{i}^{A}$ and $g^{A}$ represent the
distributions of quarks and gluons in nuclei, respectively.
$<e^{2}>$ is the average squared charge ($=\frac{2}{9}$ for light
quarks) and $C_{L,a}$ is the perturbative expansion of the
coefficient functions as it follows
\begin{eqnarray}
C_{L,a}(\alpha_{s},x)=\sum_{n=1}(\frac{\alpha_{s}}{4\pi})^{n}c^{(n)}_{L,a}(x).
\end{eqnarray}
At low $x$, the gluon contribution to the total nuclear
longitudinal structure function dominates over the singlet and
nonsinglet contributions as
\begin{equation}
F^{A}_{L}|_{x{\rightarrow}0}\simeq F^{g-A}_{L}+F^{Heavy-A}_{L}.
\end{equation}
The gluonic nuclear longitudinal structure function is given by
\begin{eqnarray}
F^{g-A}_{L}(x,Q^{2})&=&\sum_{n=1}(\frac{\alpha_{s}}{4\pi})^{n}<e^{2}>c^{(n)}_{L,g}(x){\otimes}G^{A}(x,Q^{2}).\nonumber\\
\end{eqnarray}
The nuclear heavy quark-longitudinal structure function, at low-
$x$, is depends on the nuclear gluon distribution when neglecting
the contributions due to incoming light quarks and anti-quarks in
boson gluon fusion processes. Therefore
\begin{eqnarray}
F_{L}^{Heavy-A}(x,Q^{2})&=&C_{L,g}^{Heavy}(x,Q^{2}){\otimes}G^{A}(x,Q^{2})\nonumber\\
&&{\equiv}\sum_{n=1}F^{(n),Heavy-A}_{L}(x,Q^{2}),
\end{eqnarray}
where $n$ indicates the order of $\alpha_{s}$. At low $x$ values,
Eqs.10 and 11  are explicitly dependent on the strong coupling
constant and
nuclear gluon density.\\
Similarly, in the electron-proton collision, the gluonic
longitudinal structure function is directly dependent on the gluon
distribution function. Some analytical solutions of the Altarelli-
Martinelli equations [13] using the expanding method and hard
pomeron behavior initialized by Cooper-Sarkar $\textit{et al.}$,
have been reported in last years [14-15] with considerable
phenomenological success.\\
At leading order (LO) analysis, the gluonic nuclear longetudinal
structure function is given by
\begin{eqnarray}
F^{g-A}_{L}(x,Q^{2})&=&\frac{\alpha_{s}}{4\pi}[\sum_{i=1}^{N_{f}}e_{i}^{2}]\int_{x}^{1}\frac{dy}{y}[8(x/y)^{2}(1-x/y)]\nonumber\\
&&\times G^{A}(y,Q^{2}),
\end{eqnarray}
and
\begin{eqnarray}
F_{L}^{Heavy-A}(x,Q^{2})&=&F_{L}^{c-A}(x,Q^{2})+F_{L}^{b-A}(x,Q^{2})\nonumber\\
&&+F_{L}^{t-A}(x,Q^{2}),
\end{eqnarray}
where
\begin{eqnarray}
F_{L}^{Heavy-A}(x,Q^{2},m^{2}_{H})=2e_{H}^{2}\frac{\alpha_{s}(\mu^{2}_{H})}{2\pi}\int_{xa_{H}}^{1}\frac{xdy}{y^{2}}\nonumber\\
{\times}C_{L,g}^{H}
(\frac{x}{y},\frac{m_{H}^{2}}{Q^{2}})G^{A}(y,\mu^{2}_{H}).
\end{eqnarray}
Here $a_{H}=1+4\frac{m_{H}^{2}}{Q^{2}}$, $C_{L,g}^{H}$ is the
hevay coefficient function related to the heavy quarks mass. The
scale $\mu_{H}(=\sqrt{\frac{Q^{2}}{2}+4m^{2}_{H}})$ is the mass
factorization and the renormalization scale, and
$\alpha_{s}(\mu^{2}_{H})$ is the running coupling constant. The
heavy longitudinal coefficient function can be expressed as
\begin{eqnarray}
C^{H}_{L,g}(z,\zeta)=-4z^{2}{\zeta_{H}}ln\frac{1+\beta_{H}}{1-\beta_{H}}+2{\beta_{H}}z(1-z),
\end{eqnarray}
where $\beta_{H}^{2}=1-\frac{4z\zeta_{H}}{1-z}(\zeta_{H}{\equiv}\frac{m_{H}^{2}}{Q^{2}})$.\\
The low- $x$ behavior of the nuclear gluon distribution function
in accordance with  the KLN model can be exploited to the nuclear
longitudinal structure function. Therefore $F_{L}^{A-total}$ at
low $x$ can be found as
\begin{widetext}
\begin{eqnarray}
F_{L}^{A-total}&=&\frac{\alpha_{s}}{4\pi}[\sum_{i=1}^{N_{f}}e_{i}^{2}]\int_{x}^{1}\frac{dy}{y}[8(x/y)^{2}(1-x/y)]
G^{A}(y,Q^{2})+2e_{c}^{2}\frac{\alpha_{s}(\mu^{2}_{c})}{2\pi}\int_{a_{c}x}^{1}\frac{xdy}{y^{2}}
C_{L,g}^{c}
(\frac{x}{y},\zeta_{c})G^{A}(y,\mu^{2}_{c})\nonumber\\
&&+2e_{b}^{2}\frac{\alpha_{s}(\mu^{2}_{b})}{2\pi}\int_{a_{b}x}^{1}\frac{xdy}{y^{2}}
C_{L,g}^{b}
(\frac{x}{y},\zeta_{b})G^{A}(y,\mu^{2}_{b})\nonumber\\
&&+2e_{t}^{2}\frac{\alpha_{s}(\mu^{2}_{t})}{2\pi}\int_{a_{t}x}^{1}\frac{xdy}{y^{2}}
C_{L,g}^{t} (\frac{x}{y},\zeta_{t})G^{A}(y,\mu^{2}_{t}).
\end{eqnarray}
\end{widetext}
The behavior of the ratio $R_{L}=\frac{F_{L}^{g-A}}{AF_{L}^{g-p}}$
as a function of $x$ for $Q^{2}=2,~10$ and $100~GeV^{2}$ and
nuclei $A=12$ and $A=208$ is presented in Fig.1. The transition
point between the linear and nonlinear regions is shown at low
$Q^{2}$ values in this figure. These results are comparable with
EKS [16] and EPS [17] analysis in comparison with DS [18] and HKN
[19] parameterizations at low $x$ and also with nuclear PDFs from
the LHeC perspective [20]. The magnitude of the shadowing effect
is dependent on  $Q^{2}$ values and mass number ($A$). In fact,
this model predict a large value for the shadowing effects at low
and high $Q^{2}$ values. The shadowing effects for heavy nuclei
are larger than light nuclei at a wide range of $x$ and $Q^{2}$.
Figures 2-3 show the transition point between linear and nonlinear
regions with respect to the new geometrical scaling shifted
towards very low $x$ values, and this  relates to the
nuclear mass in eA processes.\\
In Figs.4-5, we present the small-$x$ behavior of the ratio
$R_{L}^{H}$ in accordance with the traditional transition point
(Eq.5) as a function of $x$ for $Q^{2}=2,~10$ and $100~GeV^{2}$
and nuclei $A=12$ and $A=208$. In these figures, we observed
antishadowing and shadowing behaviors at low $x$ and low $Q^{2}$
values along the traditional geometrical scaling (Eq.5). In the
all cases, the depletion and enhancement in these ratios reflect
the linear(at $x>x_{c}$)/linear(at $x>x_{c}$), nonlinear(at
$x<x_{c}$)/linear(at $x>x_{c}$) and nonlinear(at
$x<x_{c}$)/nonlinear(at $x<x_{c}$) behavior for nuclei/nuclon
related to Eq.5. The enhancement in the ratio nuclei/nuclon (i.e.
antishadowing behavior) is related to the nonlinear behavior of
nuclei. This is dependence to the coherent multiple scattering
where introduces the medium size enhanced (in powers of
$A^{1/3}$)nuclear effects [22]. Indeed nuclear shadowing is
controlled by the interplay of photon fluctuations lifetime and
coherent time for transition between no shadowing and saturated
shadowing at very small $x$. The gluon shadowing is negligibly at
$x>0.01$ which covers the whole range on the NMC data [23].
 Indeed transition of nuclei from linear to nonlinear
regions is faster than transition of nucleon.\\
 The new transition points for these ratios are
shown in figures 6-7. This is consistent with Eq.6 for charm,
bottom and top quarks. Indeed the nonlinear effects are
predominant for light and heavy nuclei at low- $Q^{2}$ values. The
shadowing effects for $A=12$ are observable at $Q^{2}=2~ GeV^{2}$
and $x<10^{-6}$ and for $A=208$ are observable at $Q^{2}=2 $ and
$10~GeV^{2}$  at $x<10^{-5}$ by the charm content of the nuclei
and nucleon. For bottom and top contribution to the longitudinal
structure functions, the shadowing effects will be noticeable at
$x<10^{-7}$ and low $Q^{2}$ values, which may be expected to be
predict at the LHeC energies. A comparison between $R_{L}$(Fig.1)
and $R_{L}^{H}$(Figs.6-7) shows that transition point for going to
the shadowing region is at larger values of
$x$. This is consistent with light and heavy quarks mass.\\
\subsection{IV. Total longitudinal structure function}

Let us now discuss about the ratio of the total longitudinal
structure functions. It is well known that the inclusive
observable $F_{L}^{total}$ is strongly dependent on the gluon
distribution and heavy contributions to the structure function.
Fig.8 shows the results of the total longitudinal structure
function ratio for $A=12$ and $A=208$  at $Q^{2}$ value of
$2~GeV^{2}$, where saturation effect is more observable than other
$Q^{2}$ values. In this case, the significant nonlinear effect is
observable as this effect start to appear at
 $x{\leq}10^{-3}$ for heavy nuclei and decrease to lower $x$ values
for light nuclei. Therefore the shadowing effect For light and
heavy nuclei are observable at low $x$ values. An enhancement at
behavior of the ratio ($R_{L}^{total}$) for light nuclei (in Fig.8
on the left panel) is due to the KLN gluon model and caused by the
anti-shadowing effects. Finally the ratio of the total
longitudinal structure function decreases
as A increases.\\
\subsection{V. Summary}

In conclusion, we have observed that the KLN model for the total
longitudinal structure function ratio $R_{L}^{total}$ gives the
saturation effect of the heavy quarks effect to the light flavors
at small $x$. This ratio shows shadowing effects for heavy nuclei
at low $x$. But for light nuclei an enhancement in addition to
depletion is shown at this region. The results are close to EPS
nuclear distribution. Lastly, one important conclusion is that
heavy contribution to the total  longitudinal structure function
ratio $R_{L}^{total}={F_{L}^{A-total}}/{AF_{L}^{p-total}}$ is
considerable and cannot be neglected especially at smaller $x$ of
the LHeC
project.\\
\subsection{Acknowledgments}
We thank F.O.Dur$\widetilde{a}$es for useful discussions, comments
 and reading the manuscript.\\

\section{References}
1. LHeC workshops 2015 (http://cern.ch/lhec).\\
2.G.R.Boroun, Phys.Lett.B{\bf744} (2015)142; Phys.Lett.B{\bf741} (2015)197. \\
3. J.L.Abelleira Fernandez, et.al., [LHeC Collab.],
J.Phys.G\textbf{39} (2012)075001.\\
4. M.Klein, Ann.Phys. {\bf528},No.1-2(2016)138-144.\\
5. V.N. Gribov and L.N. Lipatov, Sov. J. Nucl. Phys.{\bf18}
(1972) 438.\\
6. L.N. Lipatov, Sov. J. Nucl. Phys.{\bf20} (1975) 93; G.
Altarelli and G. Parisi, Nucl. Phys. B{\bf126} (1977) 298; Yu.L.
Dokshitzer, Sov. Phys. JETP {\bf46} (1977) 641.\\
7. M.Praszalowicz and T.Stebel,  JHEP 04 (2013) 169.\\
8. G.Beuf and D.Royon, arXiv:hep-ph/0810.5082(2008).\\
9. T.Steble, Phys. Rev. D 88 (2013), 014026.\\
10. D.Kharzeev, E.Levin and M.Nardi, Nucl.Phys.A{\bf730}
(2004)448;Nucl.Phys.A{\bf747} (2005)609.\\
11. F.Carvalho, F.O.Dur$\widetilde{a}$es, F.S.Navarra and
S.Szpigel,
Phys.Rev.C{\bf79} (2009)035211.\\
12. E.R.Cazaroto, F.Carvalho. V.P.Goncalves and F.S.Navarra,
Phys.Lett.B{\bf669} (2008)331.\\
13. G.Altarelli and G.Martinelli, Phys.Lett.B{\bf76} (1978)89.\\
14. A.M.Cooper-Sarkar et al., Z.Phys.C{\bf39} (1998)281.\\
15. G.R.Boroun and B.Rezaei, Eur. Phys. J. C{\bf72} (2012)2221.\\
16. K.J.Eskola, V.J.Kolhinen and C.A.Salgado, Eur.Phys.J.C{\bf9} (1999)61.\\
17. K.J.Eskola, H.Paukkunen and C.A.Salgado, JHEP{\bf0807} (2008)102.\\
18. D.de Florian and R.Sassto, Phys.Rev.D{\bf69} (2004)074028.\\
19. M.Hirai, S.Kumano and T.H.Nagai, Phys.Rev.C{\bf76} (2007)
065207.\\
20. H.Paukkunen, K.J.Eskola and N.Armesto,
arXiv:hep-ph/1306.2486(2013).\\
21.E.R.Cazaroto, F.Carvalho. V.P.Goncalves and F.S.Navarra,
Phys.Lett.B{\bf671} (2009)233.\\
22.X.Guo and J.Li, Nucl.Phys.A{\bf783} (2007)587; K. Golec-Biernat et al., Nucl.Phys.B{\bf527} (1998)289.\\
23. J.Qiu and I.Vitev, Phys.Rev.Lett.{\bf 93} (2004)262301;
B.Kopeliovich et al., Phys.Rev.C {\bf 62} (2000)035204;
J.Raufeisen, arXiv:hep-ph/0204018 (2002).\\

\begin{table}
\centering \caption{Critical point $x_{c}$ along the critical line
$Q_{s}^{2}=Q^{2}$. }\label{table:table1}
\begin{minipage}{\linewidth}
\renewcommand{\thefootnote}{\thempfootnote}
\centering
\begin{tabular}{|l|c|c|c|l|} \hline\noalign{\smallskip} $Q^{2}(GeV^{2})$ & $x_{c}^{A=1}$ & $x_{c}^{A=12}$ & $x_{c}^{A=208}$ \\
\hline\noalign{\smallskip}
2 & 0.25E-5 & 0.70E-4 & 0.31E-2  \\
10 & 0.40E-8  & 0.11E-6 & 0.50E-5 \\
100 & 0.40E-12   & 0.11E-10 & 0.50E-9 \\

 \hline\noalign{\smallskip}
\end{tabular}
\end{minipage}
\end{table}
\begin{table}
\centering \caption{Critical point $x_{c}$ with new geometrical
scaling to take into account $m_{c}$. }\label{table:table1}
\begin{minipage}{\linewidth}
\renewcommand{\thefootnote}{\thempfootnote}
\centering
\begin{tabular}{|l|c|c|c|l|} \hline\noalign{\smallskip} $Q^{2}(GeV^{2})$ & $x_{c}^{A=1}$ & $x_{c}^{A=12}$ & $x_{c}^{A=208}$ \\
\hline\noalign{\smallskip}
2 & 0.41E-8 & 0.12E-6 & 0.51E-5  \\
10 & 0.50E-9  & 0.14E-7 & 0.61E-6 \\
100 & 0.31E-12   & 0.85E-11 & 0.39E-9 \\

 \hline\noalign{\smallskip}
\end{tabular}
\end{minipage}
\end{table}
\begin{table}
\centering \caption{The same Table 2 with $m_{b}$.
}\label{table:table1}
\begin{minipage}{\linewidth}
\renewcommand{\thefootnote}{\thempfootnote}
\centering
\begin{tabular}{|l|c|c|c|l|} \hline\noalign{\smallskip} $Q^{2}(GeV^{2})$ & $x_{c}^{A=1}$ & $x_{c}^{A=12}$ & $x_{c}^{A=208}$ \\
\hline\noalign{\smallskip}
2 & 0.23E-10 & 0.64E-9 & 0.29E-7  \\
10 & 0.22E-10  & 0.61E-9 & 0.27E-7 \\
100 & 0.17E-12   & 0.48E-11 & 0.21E-9 \\

 \hline\noalign{\smallskip}
\end{tabular}
\end{minipage}
\end{table}
\begin{table}
\centering \caption{The same Table 2 with $m_{t}$.
}\label{table:table1}
\begin{minipage}{\linewidth}
\renewcommand{\thefootnote}{\thempfootnote}
\centering
\begin{tabular}{|l|c|c|c|l|} \hline\noalign{\smallskip} $Q^{2}(GeV^{2})$ & $x_{c}^{A=1}$ & $x_{c}^{A=12}$ & $x_{c}^{A=208}$ \\
\hline\noalign{\smallskip}
2 & 0.81E-18 & 0.22E-16 & 0.99E-15  \\
10 & 0.38E-17  & 0.10E-15 & 0.47E-14 \\
100 & 0.20E-16   & 0.54E-15 & 0.24E-13 \\

 \hline\noalign{\smallskip}
\end{tabular}
\end{minipage}
\end{table}
\newpage
\textbf{Figure Captions}\\
Fig.1. $R_{g}$ and $R_{L}$ evaluated as a function of $x$ at
$Q^{2}=2,~ 10~ \mathrm{and}~ 100~ GeV^{2}$ for nuclei
$A=12$ and $A=208$ with the KLN model.\\

Fig.2. $R_{g}$ and $R_{L}$ evaluated as a function of the
geometrical scalings at $Q^{2}=2~ GeV^{2}$ for  nuclear
$A=12$.\\

Fig.3. The same Fig.2 for $A=208$.\\

Fig.4. The ratio $R_{L}^{H}=\frac{F_{L}^{H(A)}}{AF_{L}^{H(p)}}$
for $A=12$ at $Q^{2}=2,~ 10~ \mathrm{and}~ 100~ GeV^{2}$.\\

Fig.5. The same Fig.4 for $A=208$.\\

Fig.6. The nonlinear and shadowing behavior of  $R_{L}^{H}$
for $A=12$ in accordance with new geometrical transition point.\\

 Fig.7. The same Fig.6 for $A=208$.\\

Fig.8. $R_{L}^{total}$ for $A=12$ and $A=208$ at
$Q^{2}=2~GeV^{2}$.\\
\newpage
\begin{figure}
\centering
\includegraphics[width=1\textwidth]{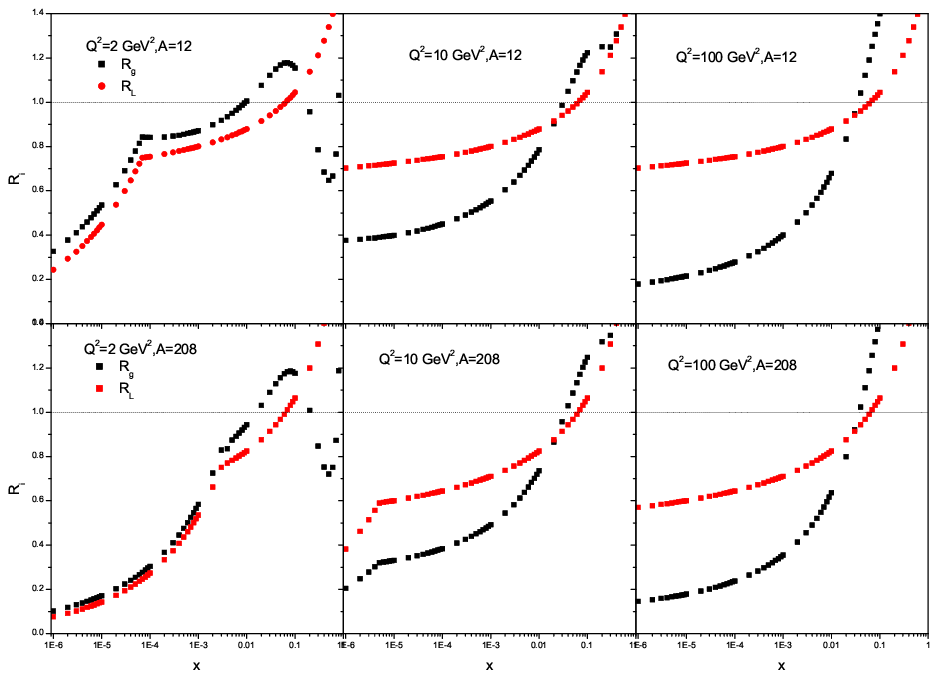}
Fig.1
\end{figure}
\begin{figure}
\centering
\includegraphics[width=1\textwidth]{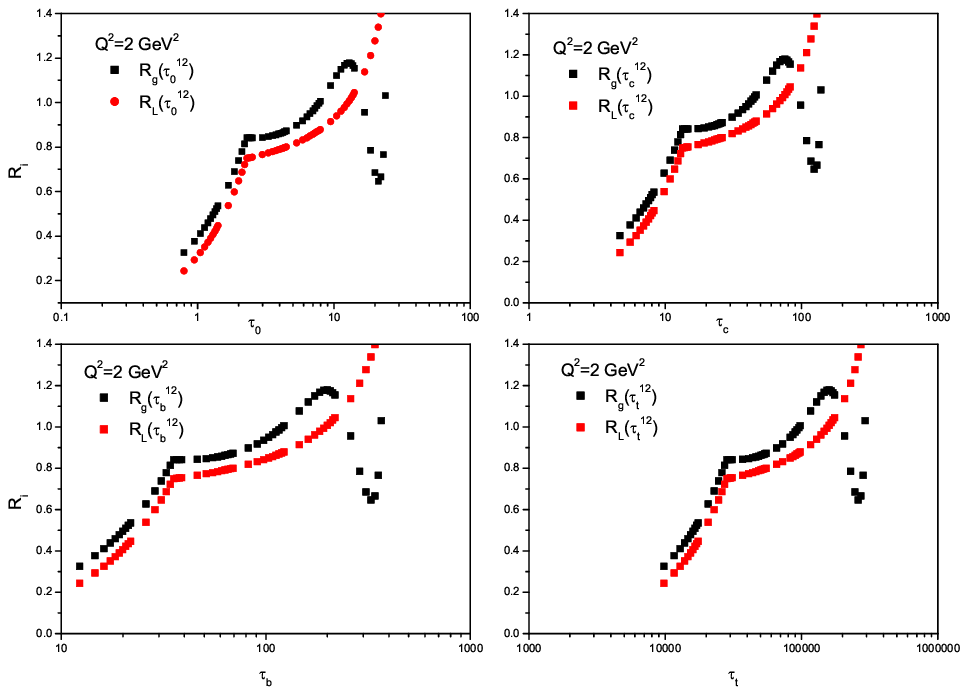}
Fig.2
\end{figure}
\begin{figure}
\centering
\includegraphics[width=1\textwidth]{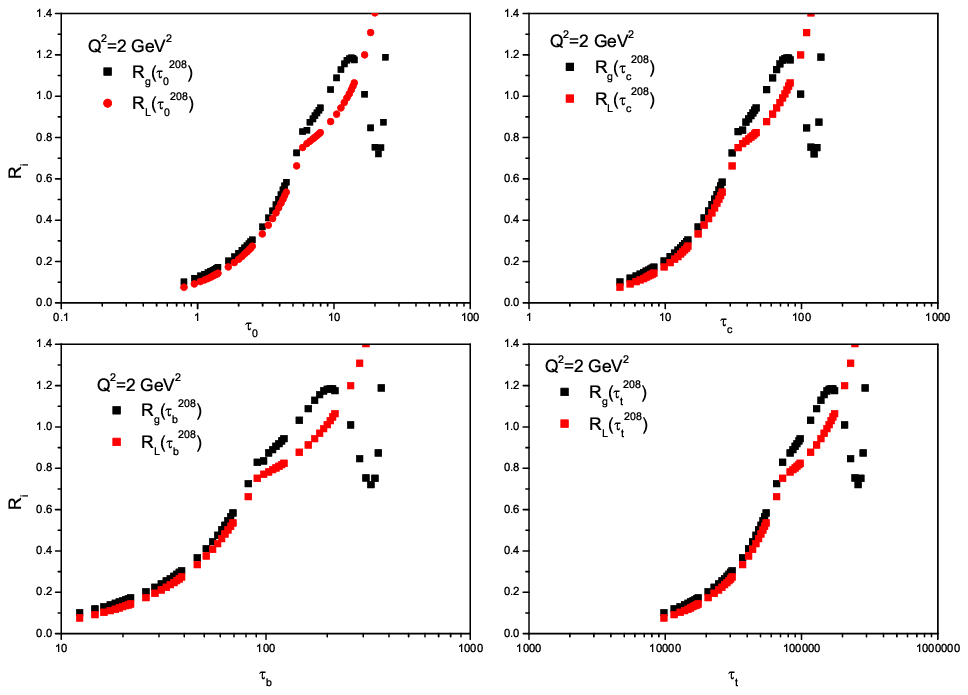}
Fig.3
\end{figure}
\begin{figure}
\centering
\includegraphics[width=1\textwidth]{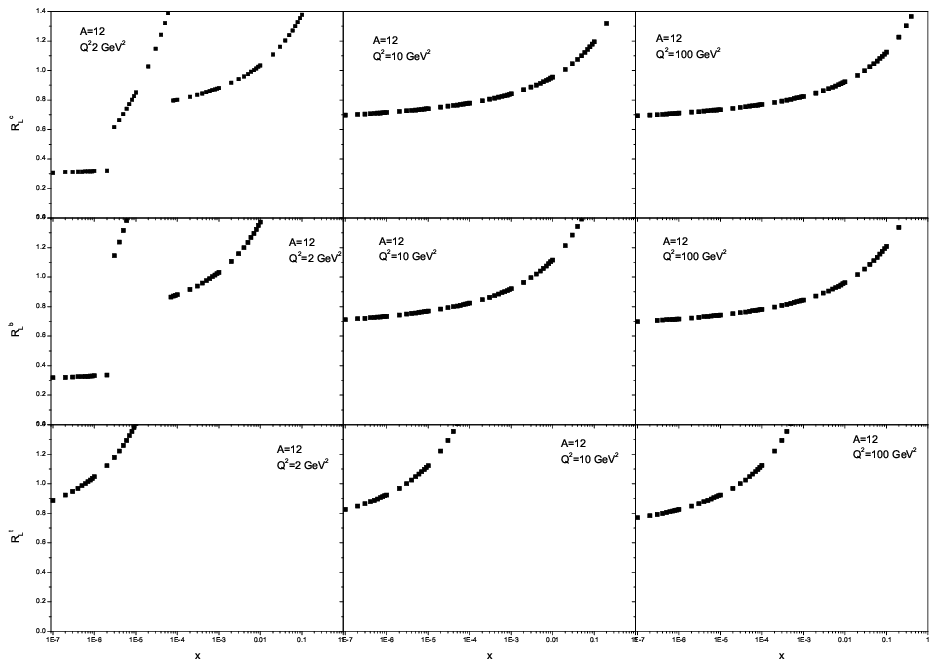}
Fig.4
\end{figure}
\begin{figure}
\centering
\includegraphics[width=1\textwidth]{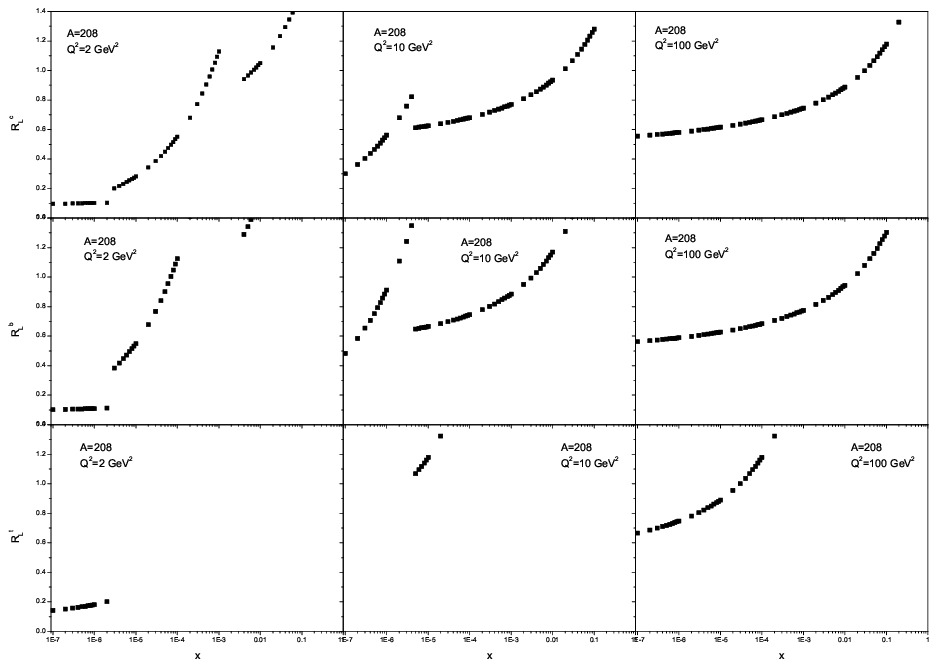}
Fig.5
\end{figure}
\begin{figure}
\centering
\includegraphics[width=1\textwidth]{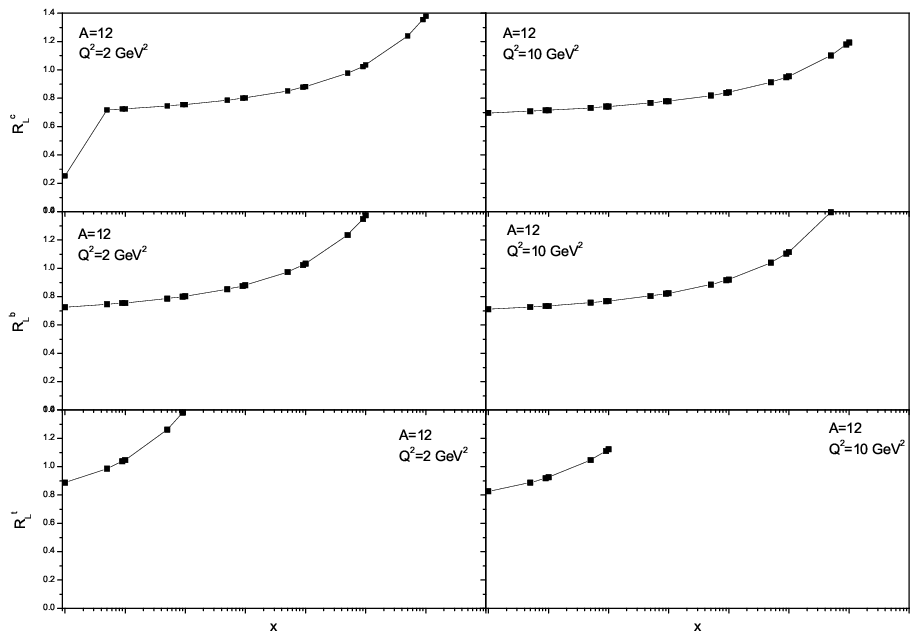}
Fig.6
\end{figure}
\begin{figure}
\includegraphics[width=1\textwidth]{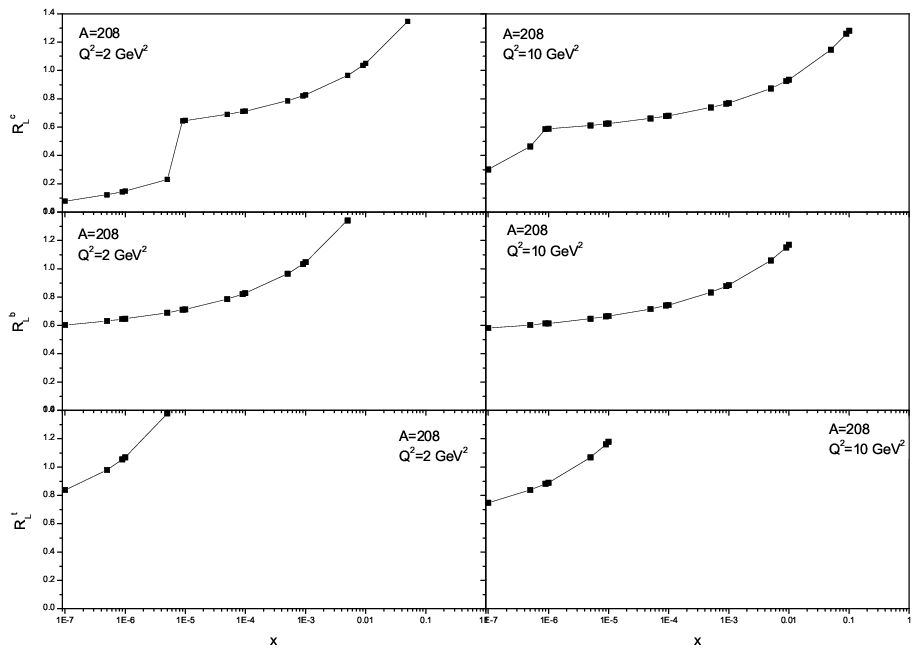}
Fig.7
\end{figure}
\begin{figure}
\centering
\includegraphics[width=1\textwidth]{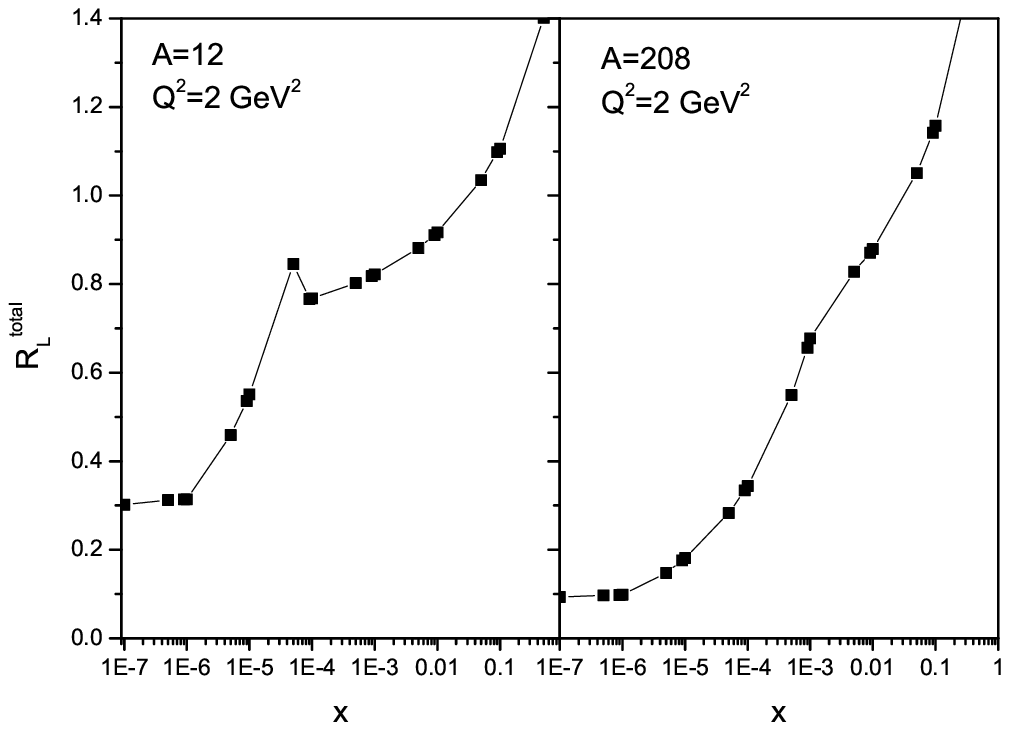}
Fig.8
\end{figure}


\end{document}